\documentclass[journal]{IEEEtran}
\usepackage[autostyle]{csquotes}
\usepackage{multirow}
\usepackage{algorithm}
\usepackage{algorithmic}
\usepackage{graphicx}
\usepackage{subfigure}
\usepackage{epstopdf}
\usepackage{xcolor}

\begin{document}
%
\title{Socially-Aware Conference Participant Recommendation with Personality Traits}
%
%
%

\author{Feng Xia,~\IEEEmembership{Senior Member,~IEEE,}
        Nana Yaw Asabere, 
        Haifeng Liu, Zhen Chen, and Wei Wang 
\thanks{The authors are with School of Software, Dalian University of Technology, Dalian 116620, China.

Corresponding author: Feng Xia; E-mail: f.xia@ieee.org}}

\markboth{IEEE Systems Journal,~Vol.~00, No.~0,~2017}%
{Shell \MakeLowercase{\textit{et al.}}: }

\maketitle

\begin{abstract}
As a result of the importance of academic collaboration at smart conferences, various researchers have utilized recommender systems to generate effective recommendations for participants. Recent research has shown that the personality traits of users can be used as innovative entities for effective recommendations. Nevertheless, subjective perceptions involving the personality of participants at smart conferences are quite rare and haven't gained much attention. Inspired by the personality and social characteristics of users, we present an algorithm called Socially and Personality Aware Recommendation of Participants (SPARP). Our recommendation methodology hybridizes the computations of similar interpersonal relationships and personality traits among participants. SPARP models the personality and social characteristic profiles of participants at a smart conference. By combining the above recommendation entities, SPARP then recommends participants to each other for effective collaborations. We evaluate SPARP using a relevant dataset. Experimental results confirm that SPARP is reliable and outperforms other state-of-the-art methods.
\end{abstract}


%
\IEEEpeerreviewmaketitle

\section{Introduction}
\IEEEPARstart{N}{owadays}, recommender systems have substantiated their necessity and importance because of how they objectively focus on solving information overload problems of users. Recommender systems provide users with personalized information services that are sometimes proactive. Due to their potential value and associated greatness in terms of research, recommender systems are studied in both academia and industry.

In the last decade, research in recommender systems have utilized two dimensional (2D) methods such as Collaborative Filtering (CF) and Content-Based Filtering (CBF) to generate recommendations for users via user profiles and items~\cite{IEEEhowto:1}. Furthermore, recommender systems research has concentrated on the performance of algorithms for recommendations and enhanced procedures of building user models to match user preferences~\cite{IEEEhowto:2}.

Within the same period, other recommender systems such as context-aware~\cite{IEEEhowto:3},~\cite{IEEEhowto:4}, hybrid~\cite{IEEEhowto:5},~\cite{IEEEhowto:6}~and socially-aware~\cite{IEEEhowto:7},~\cite{IEEEhowto:8}~have been developed in a variety of domain-specific applications. Such applications include mobile multimedia~\cite{IEEEhowto:9},~\cite{IEEEhowto:10}~and data mining~\cite{IEEEhowto:11}. While many of these recommender systems have been proposed for user modeling, little attention has been paid on analyzing the personality information involved in modeling recommendation processes~\cite{IEEEhowto:12}-\cite{IEEEhowto:14}. Nevertheless, some researchers have combined social information and personalization in their recommendation procedures. For example in~\cite{IEEEhowto:15},~the social context of documents is added as a layer to textual content to provide Personalized Social Document Representations.

The global organizations of academic conferences are very important for researchers and academicians. Conferences enable interactions and collaborations between researchers of different races and cultures. During a smart conference event, participants usually interact, socialize and introduce themselves to each other. Some participants at a conference may know each other already from the past and thus may have strong social ties~\cite{IEEEhowto:8}. Other participants who have the same research interests but do not know each other and thus have weak social ties may want to familiarize themselves with one another.

The promotion of interactions and research discussions among participants are the main aims of academic conferences. However, the rapid growth of information introduces many challenges to technology applications in different scenarios~\cite{IEEEhowto:16}. Particularly, participants at smart conferences find it difficult to deal with multiple sources of data that are constantly produced at the conference. As a result, conference participants often miss important academic and social opportunities, such as collaboration and co-authorships. In addition, it is not an easy task to find personalized information according to specific preferences and needs of users.

Recent studies on people (user) recommendation have concentrated on suggesting people the user already knows. Connecting/linking to strangers within the conferences can be valuable for participants in many ways~\cite{IEEEhowto:17}. These include: (i) getting reliable collaborative research help or advice, (ii) acquiring research opportunities that are beyond those available through existing personality and social ties~\cite{IEEEhowto:8}, (iii) discovering new routes for potential research development and (iv) learning about new research projects and assets that can be used to leverage and connect/link with subject matter experts/researchers and other influential people at the conference.

At the presentation sessions of smart conferences or the main conference venue, it is important to establish interactive mechanisms that will allow researchers who do not know each other to approach themselves. Usually a participant's personality (human behavior) determines whether he/she is approachable or not~\cite{IEEEhowto:12}-\cite{IEEEhowto:14}. Personality traits such as openness to experience, extroversion, agreeableness, conscientiousness and neuroticism are very important and should be considered in the establishment of an interactive scenario between participants at a smart conference.

Furthermore, a user's personality is critical for eliminating cold-start problems in recommender systems. In this paper, we try to enhance the interactions, collaborations and social awareness of participants of a smart conference by embedding personality as part of our recommendation procedure for collaborative participation. Our previous work~\cite{IEEEhowto:8}~involved the generation of presentation session venues for participants based on a combination of similar tagged ratings of research interests and social ties. Motivated by the personality and social characteristics of users, this paper moves a step further from our work in~\cite{IEEEhowto:8}~and proposes an algorithm called \textit{Social and Personality Aware Recommendation of Participants} (\textit{SPARP}). The main goal of \textit{SPARP} is to model the personality and social awareness of participants at their recommended presentation session venues so that further recommendations consisting of co-authorships, friendships and collaborative scenarios can be generated for participants. We suggest a novel method for recommending strangers in a smart conference with whom the shares similar personality interests but weak ties. Based on computed similarities of research interests and interpersonal relationships (more accurately predicted social ties) among participants, our method hybridizes~\cite{IEEEhowto:5},~\cite{IEEEhowto:6}~these entities to generate effective recommendations for participants.

\subsection{Contributions}
The major contributions in this paper include the following:
\begin{itemize}
\item Through the computations of Pearson correlations (personality) and estimated (accurate) social ties of participants, we develop an innovative algorithm that recommends individual participants to each other at smart conference sessions.
\item By computing the estimated (accurate) social ties of participants, we determine the extent and levels of interpersonal influence and relationships between participants, which we use in our approach to generate effective weighted hybrid (social and personality) recommendations.
\item Additionally, our proposed recommender algorithm measures the extent of personality trait relationships and similarities among participants to generate effective weighted hybrid (social and personality) recommendations.
\item Our method quantifies that even if users (participants) have low levels of tie strengths, they can still gain an effective weighted hybrid recommendation through a combination of strong similar personality traits and weak ties.
\item Our approach innovatively brings unknown/strange participants to an active participant, in contrast to the exploration and search approach, and can be viewed as a smart conference example of a social matching system.
\item We differentiate and compare our work with related/existing works to ascertain the significance of our recommendation method.
\item Finally, through a relevant dataset, our methodology is testified through experiments in order to obtain results for comparison with existing state-of-the-art methods.
\end{itemize}

\subsection{Organization}
The rest of the paper is organized as follows. Section II presents related work. Section III discusses our recommendation model, approach and algorithm. In Section IV, we discuss our experimentation/evaluation procedure and analyze the results achieved. Section V finally concludes the paper.

\section{Related Work}
A reasonable amount of research work consisting of user recommendation and linkages at academic conferences and organizations have been reported in recent years. In this section, we present some related work consisting of the following: (i) Collaborative Recommendations and Link Predictions in Academic Conferences (ii) Academic and Organizational Collaboration Recommendations and (iii) Personality-Aware Recommendations.

\subsection{Collaborative Recommendations and Link Predictions in Academic Conferences}
Social Network Analysis (SNA) has been explored in many contexts towards different goals. Various researchers such as ~\cite{IEEEhowto:18}-\cite{IEEEhowto:24}~have successfully exploited recommender systems and other relevant techniques in different social networks. Academic social networks such as conferences, symposia and workshops are organized globally to enhance knowledge through research and collaboration.

In terms of collaborative recommendations/linkages at conferences, Chin \textit{et al}.~\cite{IEEEhowto:18}~used offline proximity encounters to create a system for finding and connecting people at a conference in order to help attendees meet and connect with each other. Using relevant data, they discovered that for social selection, more proximity interactions will result in an increased probability for a person to add another as a social connection (friend, follower or exchanged contact).

Similar to~\cite{IEEEhowto:18},~Chang \textit{et al}.~\cite{IEEEhowto:19}~reported their work in Nokia Find and Connect to solve the problem of how to use mobile devices and the indoor positioning technology. Their approach was aimed to help conference participants enhance real-world interactions and improve efficiency during the conference. They used location and encounters, together with the conference basic services through a mobile User Interface (UI).

\textit{Conferator} is a novel social conference system that provides the management of social interactions and context information in ubiquitous and social environments~\cite{IEEEhowto:20}. Using RFID and social networking technology, \textit{Conferator} provides the means for effective management of personal contacts according to information pertaining to before, during and after a conference. Atzmueller \textit{et al}.~\cite{IEEEhowto:20}~described the \textit{Conferator} system and discussed analytical results of a typical conference using \textit{Conferator}.

Similar to~\cite{IEEEhowto:20},~Scholz \textit{et al}.~\cite{IEEEhowto:21}~focused on face-to-face contact networks collected at different conferences using the social conference guidance system, \textit{Conferator}. Precisely, they investigated the strength of ties and its connection to triadic closures in face-to-face proximity networks. Furthermore, they analyzed the predictability of all new and recurring links at different points of time during the conference. They also considered network dynamics for the prediction of new links during a conference.

In the same vain as~\cite{IEEEhowto:21}, Barrat \textit{et al}.~\cite{IEEEhowto:22} investigated the data collected by the Live Social Semantics application during its deployment at three major conferences, where it was used by more than 400 people. Their analyses showed the robustness of the patterns of contacts at various conferences, and the influence of various personal properties (e.g. seniority, conference attendance) on social networking patterns.

Our previous work~\cite{IEEEhowto:8},~proposed a novel venue recommender algorithm to enhance smart conference participation. Our proposed algorithm, Social Aware Recommendation of Venues and Environments (\textit{SARVE}), computes the Pearson correlation and social characteristic information of conference participants. \textit{SARVE} further incorporates the current context of both the smart conference community and participants in order to model a recommendation procedure using distributed community detection.

\subsection{Academic and Organizational Collaboration Recommendations}
In terms of academic social networks, Brandao \textit{et al}.~\cite{IEEEhowto:23}~used concepts from SNA to recommend collaborations in academic networks. They proposed two new metrics for recommending new collaborations or intensification of existing ones. Each metric considers a social principle (homophily and proximity) that is relevant within the academic context. Their focus was to verify how these metrics influence the resulting recommendations. They also proposed new metrics for evaluating the recommendations based on social concepts (novelty, diversity and coverage) that have never been used for such a goal.

In the same vain as~\cite{IEEEhowto:23},~Li \textit{et al}.~\cite{IEEEhowto:24}~satisfied the demand of collaboration recommendation through co-authorship in an academic network. They proposed a random walk model using three academic metrics as basics for recommending new collaborations. Each metric was studied through mutual paper co-authoring. Compared with other state-of-the-art approaches, experiments on DBLP dataset showed that their approach improved the precision, recall rate and coverage rate of academic collaboration recommendations.

Meo \textit{et al}.~\cite{IEEEhowto:25}~presented an in-depth analysis of the user behaviors in different Social Sharing systems. They considered three popular platforms, Flickr, Delicious and Stumble. Upon, and, by combining techniques from SNA with techniques from semantic analysis, they characterized the tagging behavior as well as the tendency to create friendship relationships of the users of these platforms. The aim of their investigation was to verify if the features and goals of a given Social Sharing system reflects on the behavior of its users and, moreover, if there exists a correlation between the social and tagging behavior of the users.

Similar to~\cite{IEEEhowto:25},~Xu \textit{et al}.~\cite{IEEEhowto:26}~created a friend recommender system using proximity encounters and meetings as physical context called Encounter Meet. They conducted a user study to examine whether physical context-based friend recommendation is better than common friends.

Guy \textit{et al}.~\cite{IEEEhowto:17}~used social media behavioral data to recommend people a user is not likely to know, but nonetheless may be interested in. Their evaluation was based on an extensive user study with 516 participants within a large enterprise and included both quantitative and qualitative results. They found out that many employees valued the recommendations, even if only one or two of nine recommendations were interesting strangers.

In the same vain as~\cite{IEEEhowto:17},~Diaby \textit{et al}.~\cite{IEEEhowto:27}~presented a content-based recommender system which suggests jobs to Facebook and LinkedIn users. A variant of their recommender system is currently used by \textit{Work4}, a San Francisco-based software company that offers Facebook recruitment solutions. The profile of a user contains two types of data: interactions data (user's own data) and social connections data (user's friends data). Furthermore the profiles of users and the description of jobs are divided into several parts called fields. Their experiments suggested that to predict the interests of users  for jobs, using fundamental similarity measures together with their interactions data collected by \textit{Work4} can be improved upon.

\subsection{Personality-Aware Recommendations}
Personality is defined as the organized and developing system within an individual that represents the collective action of that individual's major psychological subsystems~\cite{IEEEhowto:28}. Research has shown that personality is an enduring and primary factor which influences human behaviors and that there are significant connections between peoples' tastes and interests~\cite{IEEEhowto:28}. Personality is a critical factor which influences peoples' behavior and interests. There is a high potential that integrating users' personality characteristics into recommender systems could improve recommendation quality and user experience~\cite{IEEEhowto:12}-\cite{IEEEhowto:14}. People with similar personality features are more likely to have similar preferences. For example, in~\cite{IEEEhowto:29},~people with high scores in neuroticism generated more Chinese words about religion and art. The effect of personality on human behavior has been widely studied in psychology, behavioral and economics marketing~\cite{IEEEhowto:14}.

In terms of personality-aware recommendation, Gao \textit{et al}.~\cite{IEEEhowto:29}~proposed a new approach to automatically identify personality traits with social media contents in Chinese language environments. Social media content features were extracted from 1766 Sina micro blog users, and the predicting model was trained with machine learning algorithms.

Hu and Pu~\cite{IEEEhowto:12}~aimed at addressing the cold-start problem by incorporating human personality into the collaborative filtering framework. They proposed three approaches: the first approach was a recommendation method based on users' personality information alone, the second approach was based on a linear combination of both personality and rating information, and the third approach used a cascade mechanism to leverage both resources.

In Feng and Qian~\cite{IEEEhowto:13},~three social factors: personal interest, interpersonal interest similarity and interpersonal influence, were fused into a unified personalized recommendation model based on probabilistic matrix factorization. They used the interpersonal interest similarity and interpersonal influence of users to enhance the intrinsic link among features in the latent space for cold-start users.

Chen \textit{et al}.~\cite{IEEEhowto:30}~reported their ongoing research on exploring the actual impact of personality values on users' needs for recommendation diversity. Results from a preliminary user survey showed significant causal relationship from personality factors (such as conscientiousness) to the users' diversity preference (not only over the item's individual attributes but also on all attributes when they are combined).

Recio-Garcia \textit{et al}.~\cite{IEEEhowto:31}~introduced a novel method of generating recommendations to groups based on existing techniques of collaborative filtering and taking into account the group personality composition. They tested their method in the movie recommendation domain and experimentally evaluated its behavior under heterogeneous groups according to the group personality composition.

A reflection of literature suggests that embedding the personality of users in recommender systems requires more innovative research. There is therefore an open issue on how to effectively integrate the personality social factor in different recommendation models to improve the accuracy of recommender systems.

As enumerated above, the work in this paper is similar to~\cite{IEEEhowto:18}-\cite{IEEEhowto:23}~which all involved enhancing conference participation, but differs in that we use a weighted combination of social and personality characteristics of users instead of RFID tag interactions and Wi-Fi encounter algorithms. Consequently, our work focuses more on establishing physical social relationships among conference participants through their social and personality characteristics/features. We therefore seek to model and present a recommendation procedure that involves the recommendation of participants to each other at the presentation session venues recommended in~\cite{IEEEhowto:8}~based on their interpersonal relationships and personality.
\begin{figure}
\center
\includegraphics[width=8cm]{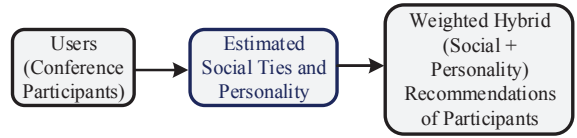}
\caption{Fundamental recommendation procedure of SPARP}
\end{figure}
Fig. 1 shows the fundamental recommendation procedure of \textit{SPARP}, which involves users, the various recommendation entities and the final weighted hybrid recommendation of participants. As shown in Fig. 2, our recommendation approach computes and hybridizes the similar personalities of participants as well as their interpersonal relationships in the form of their estimated social ties (social property) at the smart conference sessions. Additionally, we develop a recommender algorithm for discovering potential participant contacts and collaborations which can be used to establish and enhance co-authorships and friendships among participants. As a result of the enumerated differences between our work in this paper and that of other researchers, we are motivated and encouraged to embark on such a novel research issue. Furthermore, to the best of our knowledge, we are the first to tackle a recommendation research procedure that involves the combination of personality and estimated social ties at smart conference sessions.

\section{SPARP: Recommendation Model and Algorithm}
In this section, we introduce the methodology of our recommendation model. Fig. 2 illustrates our overall \textit{SPARP} recommendation model, which includes two main components, namely: interpersonal relationships and personality-based similarities of the participants. The \textit{Interpersonal Influence Analyzer} is responsible for computing the interpersonal relationships of participants through their estimated social ties. Furthermore, the \textit{Personalizer}, computes the personality profiles of participants in order to determine their personality-based similarities. As shown in Fig. 2, in our \textit{SPARP} recommendation model, there are participants in different presentation sessions who have common research interest similarities based on tagged ratings which we previously computed in~\cite{IEEEhowto:8}.
\begin{figure}
\center
\includegraphics[width=8cm]{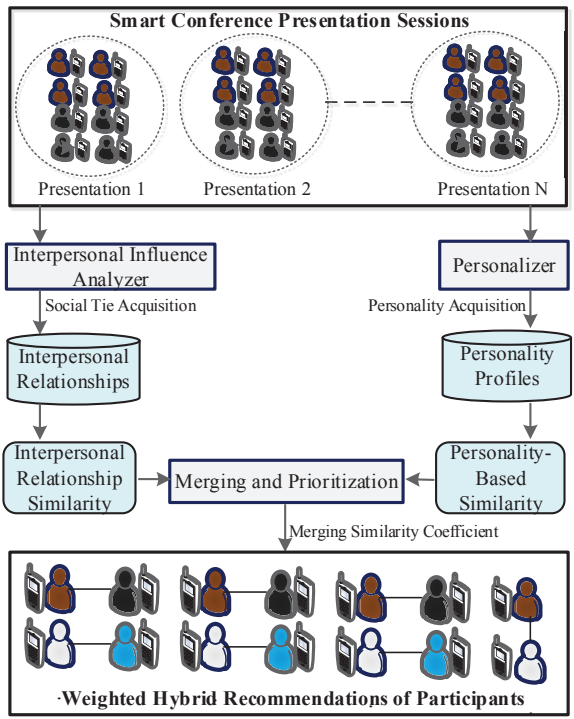}
\caption{SPARP recommendation model}
\end{figure}
The preferences of mobile device users (conference participants) can change at any time due to the changes in their surrounding environments e.g. physical conditions, location, time, their community (smart conference), etc.~\cite{IEEEhowto:32}. As a result of such changes the recommendation service in \textit{SPARP} relies on both stationary and vibrant user profiles which capture the current conference participant situation. Since \textit{SPARP} runs on mobile devices it is important that these mobile devices are equipped with the right specifications to support the recommendation service. \textit{SPARP} consequently requires standard android smartphones with relevant processing speeds (e.g. at least 1.5 GHz) and storages (e.g. 20GB Hard Disk Drive and 2GB RAM) that support the transparent usage of data involving Bluetooth, General Packet Radio Service (GPRS) and Wireless Local Area Network (WLAN).

In the first step of our \textit{SPARP} recommendation model, we extend the social ties computed in~\cite{IEEEhowto:8}~by computing a better and more accurate prediction of social ties using past and present social ties from the dataset with four different trial weight parameters. We use these weight parameters in our experiment to represent different influence proportion of the past and present social ties of participants. In the next step, \textit{SPARP} computes the similarity of personalities among participants using explicit tagged data of their personality trait ratings (1-5). Finally, in order to improve recommendation accuracy and avoid cold-start and data sparsity problems, we intuitively combine/merge the similar personalities and interpersonal relationships of participants and linearly integrate them into one merging similarity coefficient. We elaborate more on our \textit{SPARP} recommendation model and algorithm below.

\subsection{Interpersonal Relationship of Participants}
It is evident from literature that the interpersonal influence and relationships of users in a social network improves flexibility, output and efficiency. Additionally, research has also proved that social factors help improve the efficiency and accuracy of recommender systems through the avoidance and reduction of data sparsity and cold-start problems~\cite{IEEEhowto:33}-\cite{IEEEhowto:36}. A common social property which can be used to determine the interpersonal relationship of users in a social network is the computation of social ties through contact duration and contact frequency~\cite{IEEEhowto:8},~\cite{IEEEhowto:37}. Social ties are used to determine the influence two users in a network have on each other and thus the level (strong or weak) of their relationship. \textit{SPARP} utilizes the social tie property of users in a social network and computes a more accurate prediction of social ties using (1). In~\cite{IEEEhowto:8},~we computed the present social ties of participants using the product of their physical contact duration and contact frequency divided by the total time frame of the smart conference. Similarly, in this paper, through explicit data (contact duration and contact frequency) obtained from users (participants), we extend the social tie computation through a combination of past and present social ties in the dataset.

In (1), $SocTie_{a,b}(t)$ and $SocTie_{a,b}(t-\Delta t)$ are the present and past social ties between conference participants $a$ and $b$. $\beta$ is a parameter which decides the influence proportion of the present and past social ties and $\Delta t$ is the time frame used to compute the social ties between $a$ and $b$.
\begin{eqnarray}
SocTie_{a,b}(t+\Delta t)&=&\beta\times SocTie_{a,b}(t-\Delta t)+(1-\beta)\nonumber\\
& &\times SocTie_{a,b}(t)
\end{eqnarray}

\subsection{Personality of Participants}
Previous research studies on the acquisition of user personalities support the feasibility of adopting user personality information into recommender systems~\cite{IEEEhowto:12}-\cite{IEEEhowto:14},~\cite{IEEEhowto:30},~\cite{IEEEhowto:31}. Personality can be acquired through both explicit and implicit procedures~\cite{IEEEhowto:12}. Explicit procedures measure a user's personality by asking him/her to answer a list of designed and descriptive personality questions. These personality evaluation descriptors and inventories have been well recognized in the psychology field~\cite{IEEEhowto:14}. Implicit procedures acquire user information by observing the behavioral patterns of users.

In a society, people can be distinguished by their personalities. Usually people in the same personality segment are assumed to have similar behaviors or interests. Consequently, it is practical to consider that the members in a personality-based neighborhood are reliable and trustworthy recommenders to each other~\cite{IEEEhowto:12}-\cite{IEEEhowto:14}. Therefore, \textit{SPARP} employs a personality-based neighborhood approach.

The personality-based neighborhood approach is similar to that of the Pearson correlation coefficient used in recommender systems research, such as ~\cite{IEEEhowto:38},~\cite{IEEEhowto:39}. The main difference is that in the personality-based neighborhood procedure, rather than ratings, the personality traits of users are used as similarity vectors. We therefore assign a participant's personality (using explicit tagged personality ratings) in a vector similar to the procedure used in dealing with user ratings in recommender systems research. To be more exact and specific, the personality descriptor of user $a$, $P_a = (P_{a,1}, P_{a,2}, ..., P_{a,n})^T$ is an \textit{n}-dimension vector, and each dimension represents one of the characteristics in a participant's profile pertaining to one of his/her personality traits~\cite{IEEEhowto:12}.

In order to obtain reliable and standard personality descriptors for participants, we adopt the most widely and extensively used personality models within the field of psychology called the \textit{Big Five Personality Dimensions} (\textit{BFPD})~\cite{IEEEhowto:40},~shown in Fig. 3. These dimensions include the following:
\begin{itemize}
\item Openness to Experience: creative, open-minded, curious, reflective and not conventional.
\item Agreeableness: cooperative, trusting, generous, helpful, nurturing, not aggressive or cold.
\item Extroversion: assertive, amicable, outgoing, sociable, active, not reserved or shy.
\item Conscientiousness: preserving, organized and responsible.
\item Neuroticism (Emotional Stability): relaxed, self-confident, not moody, easily upset or easily stressed.
\end{itemize}
\begin{figure}
\center
\includegraphics[width=8cm]{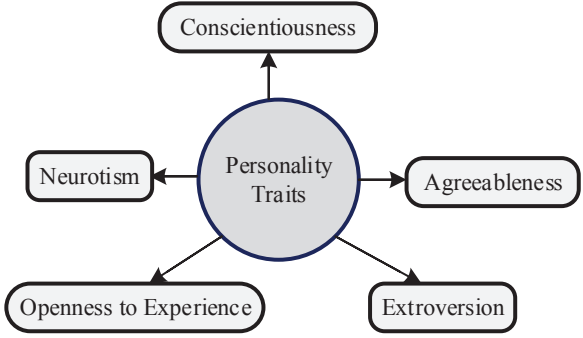}
\caption{Big five personality dimensions}
\end{figure}
\begin{equation}
Simp(a,b)=\\
\frac{\sum_{k \in K}{(p_{a,k}-\overline p_a)(p_{b,k}-\overline p_b)}}
{\sqrt{\sum_{k \in K}{(p_{a,k}-\overline p_a)^2}}\sqrt{\sum_{k \in K}{(p_{b,k}-\overline p_b)^2}}}
\end{equation}

Similar to the computation of traditional CF using Pearson correlation coefficient, we compute the personality between participants $a$ and $b$ using (2). In (2), $\overline p_a$ and $\overline p_b$ respectively denote the average of all personality trait ratings of participants $a$ and $b$. Additionally, $P_{a,k}$ and $P_{b,k}$ represent the ratings of participants $a$ and $b$ with respect to one of the personality traits $k$.
\begin{figure}
\center
\includegraphics[width=6cm]{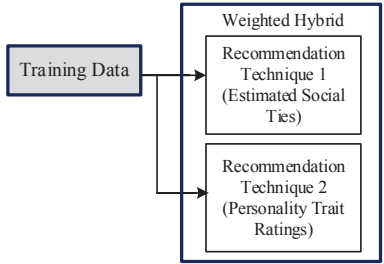}
\caption{Training phase procedure in SPARP}
\end{figure}
\subsection{Weighted (Linear) Hybrid Recommendation}
As enumerated above, we innovatively combine/merge the personality (obtained through computations of personality rating similarities) and interpersonal relationships (obtained through social tie computations) of participants. Weighted hybrids combine evidence from both recommendation techniques in a static manner, and would therefore seem to be suitable when the component recommenders have consistent relative power or accuracy across the product space~\cite{IEEEhowto:41}. Figs. 4-6 illustrate the algorithmic flow of our weighted hybrid recommender algorithm (\textit{SPARP}).
\begin{algorithm}
\caption{Pseudocode for weighted hybrid recommendation of conference participants}
\begin{algorithmic}[1]
\STATE //Declare and initialize variables
\STATE $i$, $j$ and $n$;     // Integer variables
\STATE $thresholdVal$, $pastSocialTie[n]$, $presentSocialTie[n]$, $personality[n]$ and $mergeSim[n]$;     // Floating variables
\STATE $Participants [n]$;     // Array of participants of size n
\FOR{i=0 to i\textless{n};i++}
\FOR{j=0 to j\textless{n};j++}
\STATE Compute past social ties using [$(freq*dur)/totalTime$] and store in $pastSocialTie[n]$
\STATE Compute present social ties using [$(freq*dur)/totalTime$] and store in $presentSocialTie[n]$
\STATE Calculate estimated social tie using Eq. (1) and specified $\beta$ value
\STATE Compute personality correlations using Eq. (2) and store in $personality[n]$
\STATE Merge $personality[i][j]$ with estimated $socialTie[i][j]$ and
          and store in $mergeSim[n]$
\ENDFOR
\ENDFOR
\STATE // Weighted hybrid socially-aware recommendation
\FOR{i=0 to i\textless{n};i++}
\IF{$mergeSim[i]\geq thresholdVal$}
\STATE Generate hybrid recommendation
\ENDIF
\ENDFOR
\end{algorithmic}
\end{algorithm}
\begin{figure}
\center
\includegraphics[width=8cm]{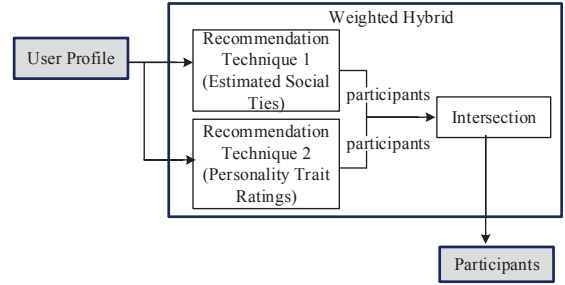}
\caption{Participant profile modeling in SPARP}
\end{figure}
Fig. 4 depicts the training phase of \textit{SPARP}, where each individual recommendation technique processes the training data. As shown in Fig. 5, after the training phase, user profiles of participants are generated for the test users. Consequently, the recommendation techniques jointly propose participants who have common intersections of user profiles in terms of social ties and personalities. Participant generation is necessary to identify those participants that will be considered in the weighted hybrid recommendation. As illustrated in Fig. 6, the participants are then sorted out through their combined weighted score and high merging similarity coefficients validates a top weighted hybrid recommendation for an active user (participant).The merging procedure shown in Fig. 6 improves the recommendation of participants who may have a combination of weak social ties (may not know each other) and high personality rating levels. To be more specific, we utilize the weighted (linear) hybrid formula below to compute the similarity between participants $a$ and $b$.
\begin{figure}
\center
\includegraphics[width=8cm]{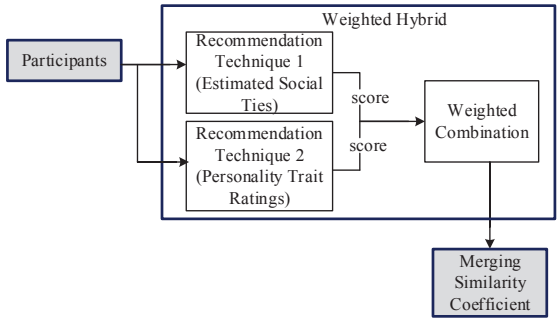}
\caption{Merging Similarity Procedure in SPARP}
\end{figure}

\begin{equation}
Sim(a,b)=SocTie_{a,b}(t+\Delta t)+Simp(a,b)
\end{equation}

Equation (3) combines the results of (1) and (2) to finally compute the similarity between $a$ and $b$ in terms of interpersonal relationships and personalities of participants. Additionally, in our experiment, we utilize $\gamma$ in (4) to set a threshold for to (3) so that we can effectively determine and generate weighted hybrid recommendations for participants.

\begin{equation}
Sim(a,b)\geq\gamma
\end{equation}

In our proposed recommender algorithm, steps 1-4 declare relevant variables, steps 5-9 compute past, present and estimated social ties of participants respectively. The similarity of the personalities of participants is computed in step 10. Step 11 merges the estimated social ties and similarity of the personalities of participants. The final steps (14-19) generate weighted hybrid recommendations for participants based on a merging similarity coefficient and threshold value.

\section{Experimentation}
In this section, we embark on a series of experiments to evaluate the performance of our proposed recommender model/algorithm (\textit{SPARP}). Initially, we introduce the compared baseline methods, then we discuss the experimental dataset and parameters. We further elaborate on the evaluation metrics employed and finally analyze the experimental results achieved.

\subsection{Baseline Methods}
To achieve effective experimental results, we compared our method to two other state-of-the-art approaches which involved enhancing social interactions and participant recommendations at conferences. These methods include the work done by Scholz \textit{et al}.~\cite{IEEEhowto:21} and Barrat \textit{et al}.~\cite{IEEEhowto:22}.

Scholz \textit{et al}.~\cite{IEEEhowto:21}~studied two aspects in the context of analyzing the contact behavior of participants at conferences. Initially, they considered the link prediction problem in evolving face-to-face contact networks. Secondly, they analyzed triadic closure at conferences using tie strengths. Specifically, they considered network dynamics for the prediction of new participant links at conferences and introduced an innovative approach of analyzing the tie strengths of conference participants and its connection to triadic closures in face-to-face proximity networks. They modeled the social network as an undirected multi-graph which involved a set of participants, an edge and a weight representing contact between two participants with a contact duration. In their dataset, more than the half of all cumulated face-to-face contacts are less than 200 seconds and the average contact duration is less than one minute, but very long contacts were also observed. We denote the method in~\cite{IEEEhowto:21}~as C1. Since C1 provides social contacts to support interaction of conference participants thereby recommending participants to each other, we compare C1 to \textit{SPARP} to verify its performance.

The Live Social Semantics (LSS) in~\cite{IEEEhowto:22}~involves a \textit{Sociopatterns} platform that enables the detection of Face-to-Face (F2F) proximity of conference participants wearing the RFID badges. The LSS architecture registers the contact events taking place within the range of RFID readers. The data of contacts is stored as a network, which allows the establishment of aggregated contact networks at the conference as follows: nodes represent individuals, and an edge is drawn between two nodes if at least one contact event took place between the corresponding conference participants. Each edge is weighted by the number of contact events or the total duration spent in F2F proximity. For each node, its degree (number of neighbors on the network) gives the number of different conference participants with whom the user has been in contact, and the strength (sum of the weights of the links) is defined by the total time this person spent in F2F interaction with other conference participants. We denote the method in~\cite{IEEEhowto:22}~as C2. LSS uses contact duration and contact frequency to determine the tie strength of conference participants. This is done to establish and recommend participants to each other. Due to the similar approach of C2 and \textit{SPARP}, we conduct a methodological comparison to substantiate the performance of our method.

In our experiment, we particularly try to answer the following questions:

\begin{itemize}
\item In terms of the utilized evaluation metrics, what is the overall performance of \textit{SPARP} in comparison to the other methods?
\item What is the impact of $\beta$ in \textit{SPARP} in terms of lower and higher levels of accuracy?
\item What is the effect of cold-start and data sparsity in \textit{SPARP}?
\end{itemize}

\subsection{Dataset and Experimental Parameters}
We utilized the ICWL 2012 dataset from our previous work~\cite{IEEEhowto:8}. We gathered new social tie data from the same 78 users in ~\cite{IEEEhowto:8}~and categorized it as present social ties and used the previous social ties data as the past social ties of users (participants). Both social tie data (past and present) have a total time frame of 12 hours (720 minutes). Additionally, as shown in Fig. 7, the highest contact durations and frequencies (times of contact) for both social tie data are 80 minutes and 7 respectively. Furthermore, we gathered explicit personality data from the same users which involved personality trait ratings of 1-5 using the \textit{BFPD}. This enabled us to use (2) to compute the similarity of personalities of  participants in the dataset. As shown in Fig. 7 and Table I, our dataset mainly comprises of past and present social ties data as well as personality data.
\begin{figure}
\centering
\subfigure[Past social ties]{
\label{fig:subfig:a} 
\includegraphics[width=4cm]{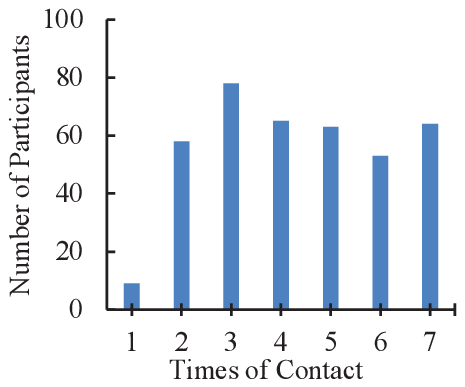}}
\subfigure[Present social ties]{
\label{fig:subfig:b} 
\includegraphics[width=4cm]{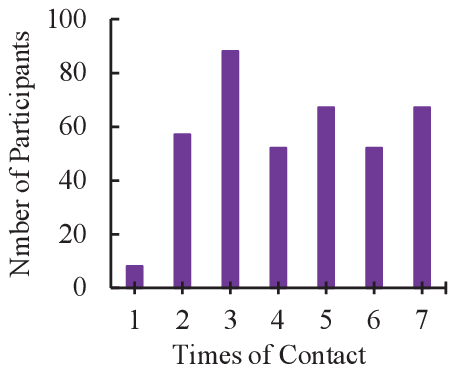}}
\caption{Contact frequency trends}
\label{fig:subfig} 
\end{figure}
\begin{table}
\centering
\caption{Personality Trait Rating Trends of Participants}
\label{table_1}
\begin{tabular}{c c c c c c}
\hline
\hline
 & \multicolumn{5}{c}{Ratings Levels of Participants} \\
\hline
Personality Traits      & 1     & 2	    & 3	    & 4	    & 5 \\
\hline
Openness to Experience	& 9     & 13	& 27    & 16	& 12 \\
Extroversion	        & 8     & 14	& 17	& 19	& 19 \\
Agreeableness           & 12	& 18	& 14	& 18	& 15 \\
Conscientiousness       & 10	& 12	& 23	& 19	& 13 \\
Neuroticism             & 13	& 18	& 16	& 19	& 11 \\
 \hline
\hline
\end{tabular}
\end{table}

Fig 7(a) and Fig. 7(b) respectively illustrate the contact frequency trends for past and present social ties. The contact frequency trends in Fig. 7 show the times of contact against the number of participants (i.e. the number of participants and their respective times of contact). Furthermore, Fig. 8 depicts the contact duration trends for past social ties between participants in minutes. For example, referring to Fig. 8, 44 participants had a contact duration of 5 minutes. Additionally, Fig. 9 depicts the contact duration trends for present social ties between participants in minutes. For instance, referring to Fig. 9, 27 participants had a contact duration of 80 minutes. We divided the dataset into training and test sets representing 70\% and 30\% respectively.

The computations of the merging similarity coefficients ranged from 0.1 to 1.0. We therefore used merging similarity coefficients ranging between 0.5 and 1.0 for testing and the rest of the computed data for training. We observed that weighted hybrid recommendations were more successful for participants whose merging coefficient similarities fell between 0.8 and 1.0. We therefore used this range as the threshold for prediction quality in accordance to the dataset.

\subsection{Metrics}
In order to evaluate our proposed recommender algorithm and compare its performance with the other state-of-the-art methods (C1 and C2), we focused on prediction quality and utilized three relevant evaluation metrics to accomplish this task. The evaluation metrics we utilized include: \textit{Accuracy}, \textit{Mean Absolute Error} (\textit{MAE}) and \textit{Normalized MAE} (\textit{NMAE}). We chose these metrics to maintain consistency and uniformity with most previous research that involved the utilizations of such metrics.
\begin{figure}
\center
\includegraphics[width=8cm]{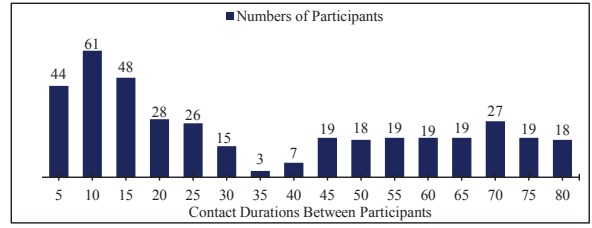}
\caption{Contact duration trends - past social ties}
\end{figure}

\begin{figure}
\center
\includegraphics[width=8cm]{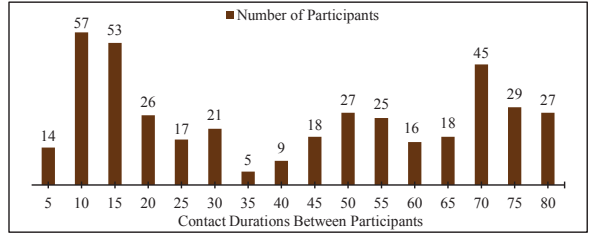}
\caption{Contact duration trends - present social ties}
\end{figure}
Accuracy metrics measure the quality of nearness to the truth or true value achieved by the recommender system/algorithm. Accuracy is the most well-known and used metric in the field of Artificial Intelligence (AI). In recommender systems research, Accuracy metrics is formulated as shown in (5)~\cite{IEEEhowto:42}.
\begin{equation}
Accuracy=\frac{\textit{number of successful recommendations}}{\textit{number of recommendations}}
\end{equation}

As depicted in (5), we assume that a \enquote{\textit{successful recommendation}} is equivalent to how useful the recommended item (participant) is and its closeness to the user's real interests.
\begin{equation}
MAE= 1-Accuracy
\end{equation}

MAE is a prediction accuracy metrics that measures the absolute deviation between each predicted rating and each user's real rating of an item. Due to the fact that both Accuracy and MAE utilize binary functions, it can be considered and assumed that the (MAE) number of recommender predictions is equal to the (accuracy) number of recommendations~\cite{IEEEhowto:42}. Consequently, as elaborated by Olmo and Gaudioso~\cite{IEEEhowto:42}, Accuracy and MAE can be reformulated using (6), which indicates that a lower MAE means better prediction performance of a recommender algorithm/system.
\begin{equation}
NMAE=\frac{MAE}{r_{max}-r_{min}}
\end{equation}

Due to the fact that different recommender systems/algorithms may use different numerical scales, we utilized \textit{NMAE} in our experiment so that experimental errors can be expressed on a full normalized scale. We therefore used (7) to compute \textit{NMAE}. In (7), $r_{max}$ and  $r_{min}$  are the upper and lower bounds of user personality trait ratings respectively in the dataset. Therefore, in accordance to the dataset, $r_{max}$=5 and $r_{min}$=1.
\begin{figure}
\centering
\subfigure[Accuracy performance]{
\label{fig:subfig:a}
\includegraphics[width=8cm]{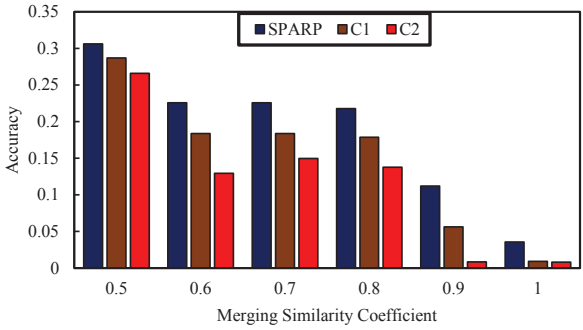}}
\subfigure[MAE Performance]{
\label{fig:subfig:b}
\includegraphics[width=8cm]{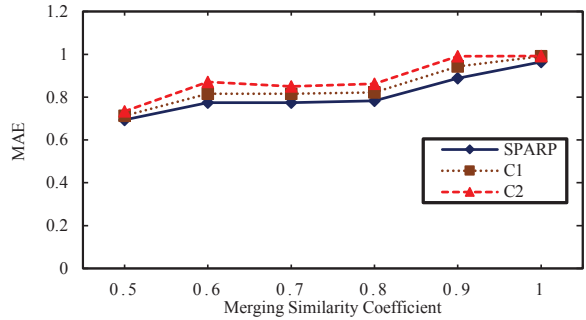}}
\caption{Weighted hybrid recommendation based on $\beta =0.1$}
\label{fig:subfig}
\end{figure}

\begin{figure}
\centering
\subfigure[Accuracy performance]{
\label{fig:subfig:a}
\includegraphics[width=8cm]{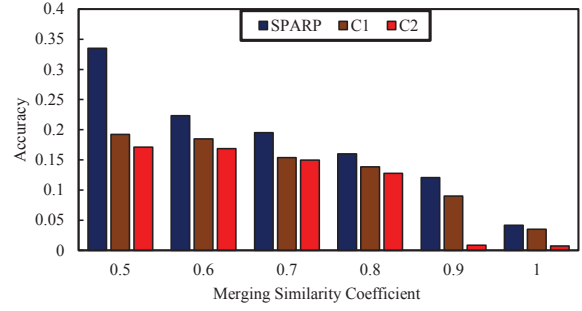}}
\subfigure[MAE Performance]{
\label{fig:subfig:b}
\includegraphics[width=8cm]{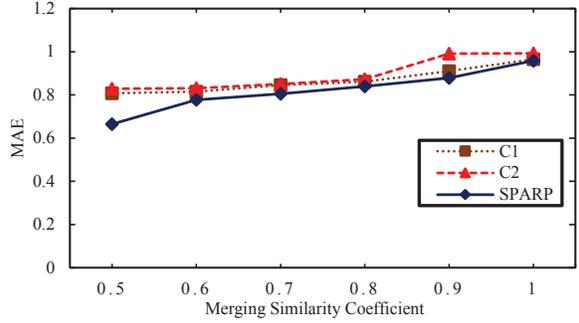}}
\caption{Weighted hybrid recommendation based on $\beta =0.2$}
\label{fig:subfig}
\end{figure}

\subsection{Experimental Results and Analysis}
As elaborated above, our experiment aimed to initially analyze the accuracy of our weighted hybrid recommendation method which combines social awareness and personality of participants. Based on similarity computations involving social information and personality, we further computed the accuracies and subsequent MAEs for each recommendation method using different weight parameters ($\beta$=0.1, 0.2, 0.3 and 0.4).

In terms of accuracy, the experimental results for \textit{SPARP} are more accurate and exact especially at higher recommendation merger values in accordance to the dataset. Referring to Fig. 10(a), where $\beta$=0.1, at the highest merging similarity coefficient (1.0), \textit{SPARP} achieved a higher accuracy (0.036) in comparison to that of C1 (0.009) and C2 (0.008). Similarly, in Fig. 11(a), where $\beta$=0.2, at the highest merging similarity coefficient (1.0), \textit{SPARP} achieved a higher accuracy (0.042) in comparison to that of C1 (0.035) and C2 (0.007). In the same vain, both Figs. 12(a) and 13(a) illustrate the effectiveness of our \textit{SPARP} method in terms accuracy and how it outperforms the other methods. These results in our experiment substantiates the fact that, in comparison to C1 and C2, \textit{SPARP} shows the ability to display and recommend more useful participants/contacts.

In terms of MAE, the experimental results for \textit{SPARP} attained lower values which corroborated better performance in comparison to the other methods. Referring to Fig 10(b), where $\beta$=0.1, at the highest merging similarity coefficient (1.0), \textit{SPARP} attained the lowest MAE value of 0.964 in comparison to C1 (0.991) and C2 (0.992). Similarly, in Fig. 11(b), where $\beta$=0.2, at the highest merging similarity coefficient (1.0), \textit{SPARP} achieved the lowest MAE (0.958) in comparison to that of C1 (0.965) and C2 (0.993). Subsequent results of MAE in Figs. 12(b) and 13(b) further corroborate the effectiveness of \textit{SPARP} in comparison to the other methods (C1 and C2). Table II summarizes the results of MAE and NMAE for the threshold merging similarity coefficients in our experiment. In Table II, lower MAE and NMAE values signify better performance. Referring to Table II, it is evident that C1 outperforms C2 and \textit{SPARP} outperforms C1. For example, in the first row of Table II, the Merging Similarity Coefficient, 0.8 ($\beta$=0.1), shows that \textit{SPARP} achieves an MAE of 0.782 which is less in comparison to that of C1 (0.821) and C2 (0.862). For the same Merging Similarity Coefficient, the NMAE of \textit{SPARP} is 0.196 which is less in comparison to that of C1 (0.205) and C2 (0.216). Our experimental results confirm that \textit{SPARP} performs better than other methods under the utilized weight parameters in terms of accuracy, MAE and NMAE. The outperformance of \textit{SPARP} implies that innovative combination of social awareness and personality traits can gain meaningful knowledge from user and user clusters in social networks to achieve effective recommendation accuracy.
\begin{figure}
\centering
\subfigure[Accuracy performance]{
\label{fig:subfig:a}
\includegraphics[width=8cm]{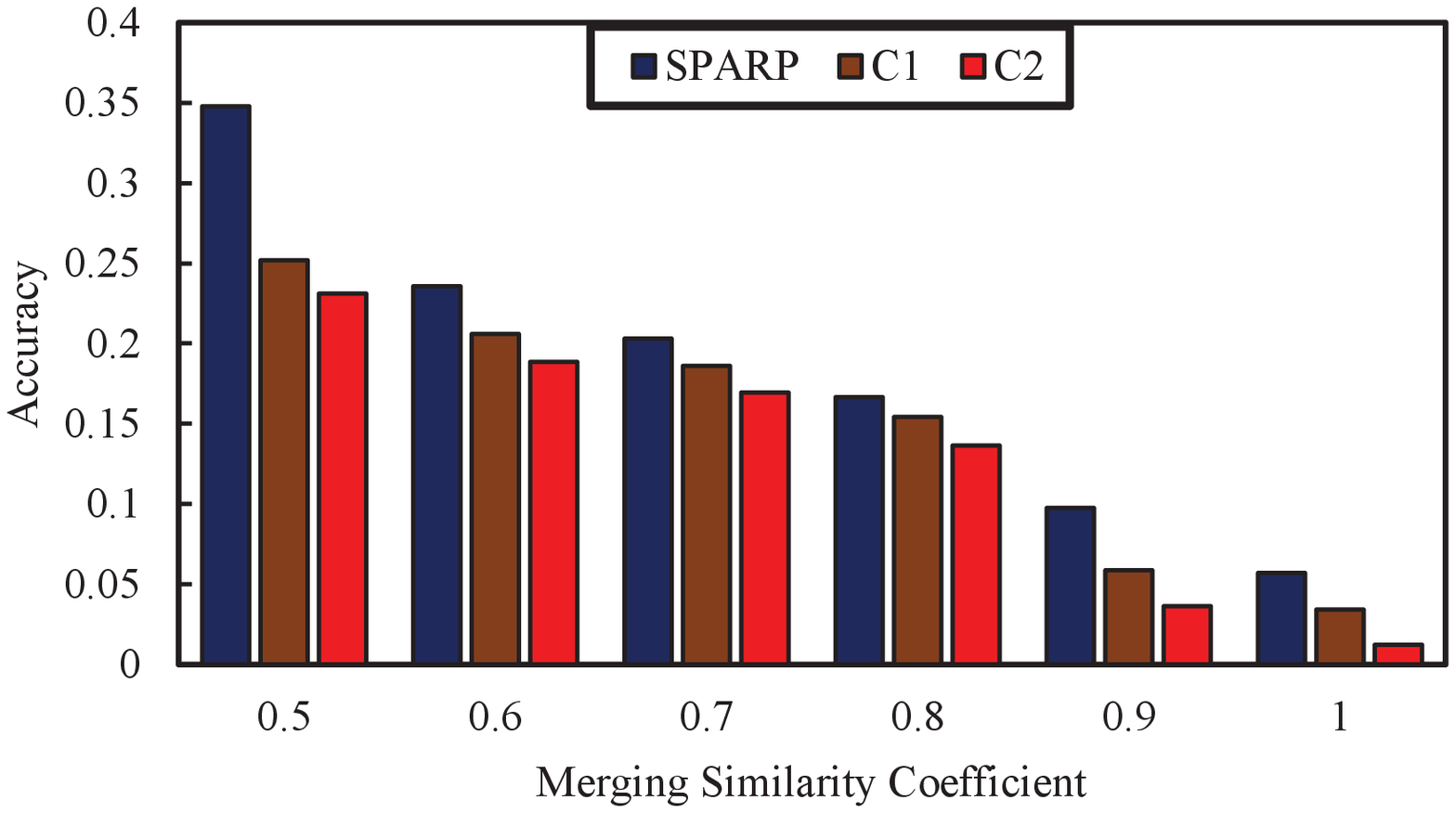}}
\subfigure[MAE Performance]{
\label{fig:subfig:b}
\includegraphics[width=8cm]{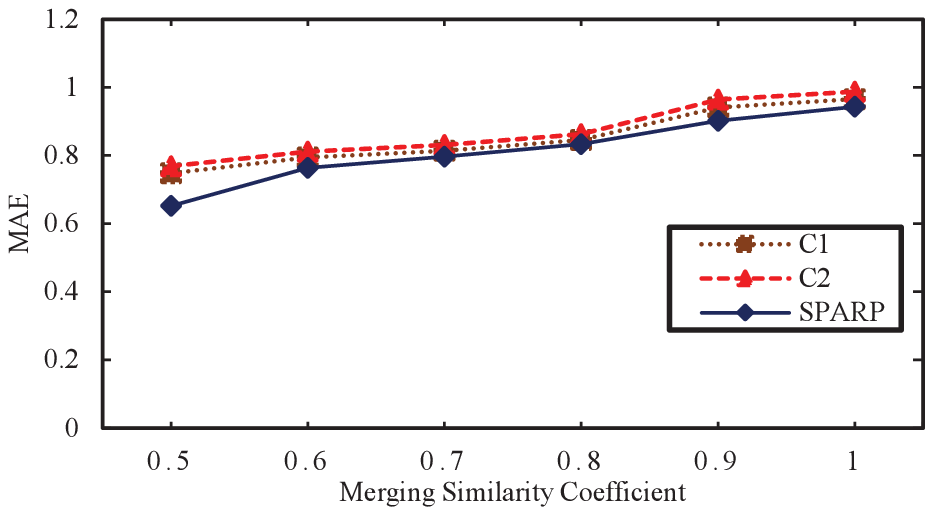}}
\caption{Weighted hybrid recommendation based on $\beta =0.3$}
\label{fig:subfig}
\end{figure}

In our experiment, we observed that even if participants had weak social ties, a strong similarity of the personality traits resulted in an effective social recommendation. We also verified that through the different weight parameters ($\beta$), the results achieved in terms of the utilized metrics were favorable. Our experimental results also depict that the different weight parameters were consistent with each of the metrics we utilized and that in each parameter, \textit{SPARP} outperformed C1 and C2.

Furthermore, referring to Figs. 10(a), 11(a), 12(a) and 13(a), at the highest merging similarity coefficient of 1.0, when $\beta$ respectively increases from 0.1 to 0.2, the accuracy of \textit{SPARP} initially upsurges from 0.035 in Fig. 10(a) to 0.042 in Fig 11(a) and further increases to 0.057 in Fig. 12(a) at $\beta$=0.3. From 0.057, the accuracy of \textit{SPARP}, increases to 0.059 at $\beta$=0.4 in Fig. 13(a). This means \textit{SPARP} attains higher accuracy levels when $\beta$ increases and we can therefore conclude that higher influence (weight) proportions of participants improves the recommendation accuracy. Correspondingly, as shown in Table II, at the highest Merging Similarity Coefficient of 1.0, the MAE of \textit{SPARP} at $\beta$=0.4 is 0.940, which is the lowest in comparison to $\beta$=0.3 (0.943), $\beta$=0.2 (0.958) and $\beta$=0.1 (0.964). Therefore, our experimental results shows that an increase in accuracy corresponds to a reduction in errors (MAE and NMAE).

\begin{table}
\centering
\caption{MAE and NMAE Performance Comparisons over the Dataset}
\label{table_2}
\begin{tabular}{c || c | c | c || c | c | c}
\hline
\hline
 & \multicolumn{3}{c||}{MAE Performance}& \multicolumn{3}{c}{NMAE Performance} \\
\hline
\multirow{3}{0.5in}{Merging Similarity Coefficient}     & \multirow{3}*{C1}     & \multirow{3}*{$SPARP$}	
& \multirow{3}*{C2}	    & \multirow{3}*{C1}     & \multirow{3}*{$SPARP$}	& \multirow{3}*{C2}	 \\
&&&&&&\\
&&&&&&\\
\hline
0.8 ($\beta$=0.1)	&0.821	&\textcolor{blue}{0.782}&0.862	&0.205	&\textcolor{blue}{0.196}	&0.216 \\ \hline
0.9 ($\beta$=0.1)	&0.944	&\textcolor{blue}{0.888}&0.991	&0.236	&\textcolor{blue}{0.222}	&0.248 \\ \hline
1.0 ($\beta$=0.1)	&0.991	&\textcolor{blue}{0.964}&0.992	&0.248	&\textcolor{blue}{0.241}	&0.248 \\ \hline
0.8 ($\beta$=0.2)	&0.862	&\textcolor{blue}{0.84}	&0.872	&0.215	&\textcolor{blue}{0.21}	&0.218 \\ \hline
0.9 ($\beta$=0.2)	&0.91	&\textcolor{blue}{0.879}&0.991	&0.228	&\textcolor{blue}{0.219}	&0.247 \\ \hline
1.0 ($\beta$=0.2)	&0.965	&\textcolor{blue}{0.958}&0.993	&0.241	&\textcolor{blue}{0.239}	&0.248 \\ \hline
0.8 ($\beta$=0.3)	&0.846	&\textcolor{blue}{0.833}&0.863	&0.211	&\textcolor{blue}{0.208}	&0.216 \\ \hline
0.9 ($\beta$=0.3)	&0.941	&\textcolor{blue}{0.902}&0.964	&0.235	&\textcolor{blue}{0.226}	&0.241 \\ \hline
1.0 ($\beta$=0.3)	&0.966	&\textcolor{blue}{0.943}&0.987	&0.241	&\textcolor{blue}{0.236}	&0.247 \\ \hline
0.8 ($\beta$=0.4)	&0.923	&\textcolor{blue}{0.845}&0.953	&0.230	&\textcolor{blue}{0.211}	&0.238 \\ \hline
0.9 ($\beta$=0.4)	&0.923  &\textcolor{blue}{0.863}&0.937	&0.231	&\textcolor{blue}{0.216}	&0.234 \\ \hline
1.0 ($\beta$=0.4)	&0.966	&\textcolor{blue}{0.940}&0.980	&0.242	&\textcolor{blue}{0.235}	&0.245 \\ \hline
\hline
\end{tabular}
\end{table}

Additionally, our experimental results exactly fit the fact that like-minded users with similar personality and social tie features are more likely to have similar interests that substantiate recommendation accuracy. Moreover, because of the effective combination of interpersonal relationships with personality, our proposed recommendation method substantially avoided cold-start problems enabling more effective social recommendations to be generated for most of the participants, in comparison to the other methods. In summary, compared with C1 and C2, \textit{SPARP}  has the minimal variation in its recommendation accuracy. This shows that \textit{SPARP} is more robust than the other methods in handling the data sparsity. Furthermore, \textit{SPARP} also exemplifies an attractive characteristic that it attains high levels of accuracy even if in a small training set. Therefore, \textit{SPARP} may be tested over a medium size subset of the original user-user matrix, which saves lots of time in an experiment.
\begin{figure}
\centering
\subfigure[Accuracy performance]{
\label{fig:subfig:a}
\includegraphics[width=8cm]{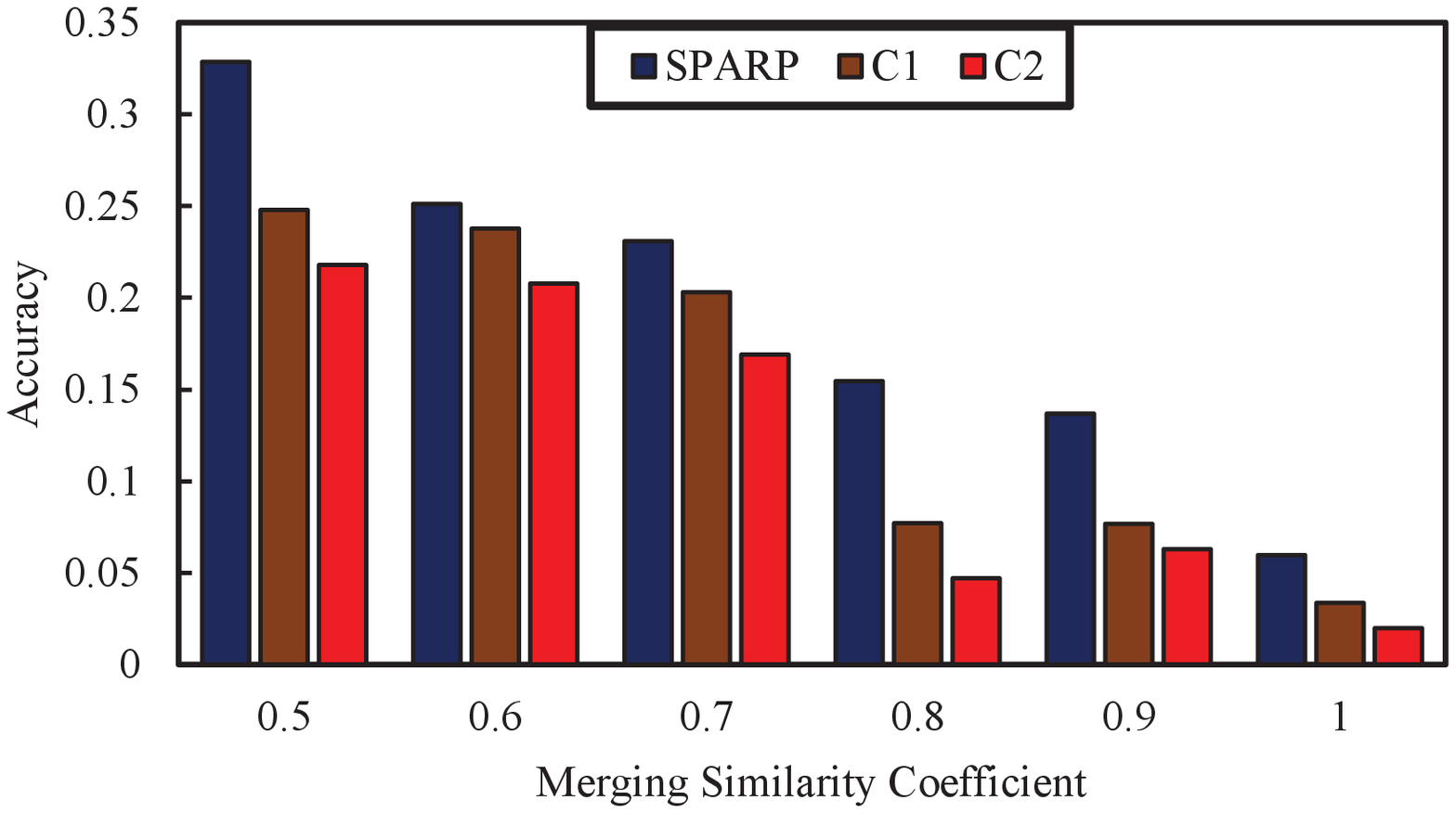}}
\subfigure[MAE Performance]{
\label{fig:subfig:b}
\includegraphics[width=8cm]{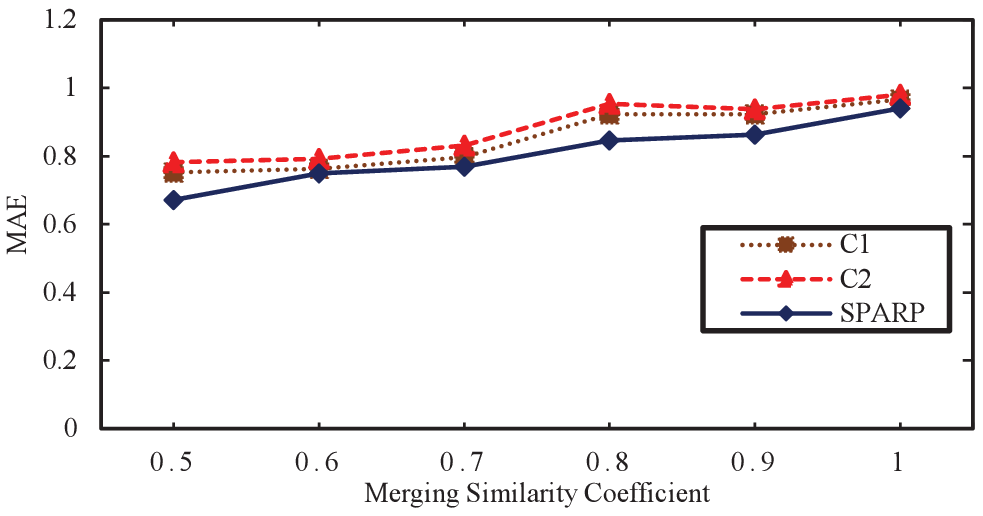}}
\caption{Weighted hybrid recommendation based on $\beta =0.4$}
\label{fig:subfig}
\end{figure}

\section{Conclusion}
In this paper, a personalized recommendation model was proposed by utilizing an algorithm (\textit{SPARP}) that combines the interpersonal relationships and personality similarities of conference participants. Specifically, through a relevant dataset which involved both past and present social tie data as well as personality data, we were able to compute a more accurate prediction of social ties among participants which enabled us to determine the extent of their interpersonal relationships. The interpersonal relationships of participants were then combined with their similar personalities (obtained through their personality trait ratings). By merging the above computations using different parameters in our experiment, we obtained weighted hybrid recommendation results that outperformed other state-of-the-art methods and were more accurate and applicable. Additionally, our algorithm reduced cold-start and data sparsity problems because of our innovative recommendation entities and hybridization procedure.

Presently, our \textit{SPARP} recommendation model is in an initial phase and only takes a user's personality traits and interpersonal relationship (estimated social ties) of the social network into consideration. As a future work, we would like to explore and utilize more social properties such as closeness centrality and selfishness in order to analyze their possible combinations with personality. Such future innovative procedures will improve weighted hybrid recommendations that involve personality and social awareness.

\ifCLASSOPTIONcaptionsoff
  \newpage
\fi

%
%
%
%
%




\end{document}